\documentclass[prd,a4paper,showpacs,preprint,byrevtex]{revtex4}
\usepackage{graphicx}
\usepackage{dcolumn}
\usepackage{amsmath}
\usepackage{array}
\usepackage{bm}
\usepackage{textcomp}

\begin{document}

\title{String deformations induced by retardation effects}

\author{Fabien \surname{Buisseret}}
\thanks{FNRS Research Fellow}
\email[E-mail: ]{fabien.buisseret@umh.ac.be}
\author{Vincent \surname{Mathieu}}
\thanks{IISN Scientific Research Worker}
\email[E-mail: ]{vincent.mathieu@umh.ac.be}
\author{Claude \surname{Semay}}
\thanks{FNRS Research Associate}
\email[E-mail: ]{claude.semay@umh.ac.be}
\affiliation{Groupe de Physique Nucl\'{e}aire Th\'{e}orique,
Universit\'{e} de Mons-Hainaut,
Acad\'{e}mie universitaire Wallonie-Bruxelles,
Place du Parc 20, BE-7000 Mons, Belgium}

\date{\today}

\begin{abstract}
The rotating string model is an effective model of mesons, in which the
quark and the antiquark are linked by a straight string. We previously
developed a new framework to include the retardation effects in the
rotating string model, but the string was still kept straight. We now go
a step further and show that the retardation effects cause a small
deviation of the string from the straight line. We first give general
arguments constraining the string shape. Then, we find analytical and
numerical solutions for the string deformation induced by retardation
effects. We finally discuss the influence of the curved string on the
energy spectrum of the model.
\end{abstract}


\maketitle

\section{Introduction}

The retardation effect between two interacting particles is a
relativistic phenomenon, due to the finiteness of the interaction speed.
Light mesons are typical systems in which these effects can
significantly contribute to the dynamics, since the light quarks can
move at a speed close to the speed of light. We developed in
Ref.~\cite{Buis} a generalization of the rotating string model (RSM)
\cite{dubi94,guba94} in order to take into account the retardation
effects in the mesons. The RSM is an effective model derived from the
QCD Lagrangian, describing a meson by a quark and an antiquark linked by
a straight string. Both particles are considered as spinless because
spin
interactions are sufficiently small to be added in perturbation. It has
been shown that the RSM was classically equivalent to the relativistic
flux tube model \cite{sema04,buis042}. This last model, firstly
presented in Refs.~\cite{laco89,tf_2}, yields meson spectra in good
agreement with the experimental data \cite{tf_Semay}. Our method to
treat the retardation effects relies
on the hypothesis that the relative time between the quark and the
antiquark must have a nonzero value. Consequently, in our approach, the
evolution parameter of the system is not the common proper time of the
quark, the antiquark and the string, but the time coordinate of the
center of mass which plays the role of an ``average" time.

We showed in Ref.~\cite{Buis} that, in the special case where the quark
and the antiquark have the same mass, the part of the total
Hamiltonian containing the retardation terms could be treated as a
perturbation. This perturbation is a harmonic oscillator in the relative
time variable, with an effective reduced mass and an effective restoring
force both depending on eigenstates of the unperturbed Hamiltonian
(which is independent of the relative time). The fundamental state of
this oscillator gives the contribution of the retardation to the masses
as well as the relative time part of the wave function. So, the relative
time wave function is a gaussian function centered around zero. This
point confirms the validity of the usual non retarded model in first
approximation. It is worth mentioning that our retardation term does not
destroy the Regge trajectories, the linear relation between the square
mass and the spin of the light mesons. Our generalized RSM
also allows to reproduce the experimental meson spectrum with a good
agreement.

This previous work must be considered as a first trial to compute an
estimation of the retardation effects in mesons. Some hypothesis were
made in order to keep the calculations workable.
In particular, the straight line ansatz was used to describe the string.
Although it simplifies the calculations, it is worth noting that the use
of a non vanishing relative time is not really compatible with a
straight string. We already gave in Ref.~\cite{Buis} a crude estimation
of the possible bending of the string due to retardation effects, and
our result was compatible with a small deviation from the straight line.
However, this point deserves a further study to confirm the validity of
our approach. Moreover, these deformations are of intrinsic interest,
since they can exist independently of the retardation effects.

Our paper is organized as follows. In Sec.~\ref{model}, we briefly
recall the approach developed in Ref.~\cite{Buis}, the RSM with a
nonzero relative time. Then, assuming that the string can be curved, we
give arguments constraining its possible shape in Sec.~\ref{cons}, and
we make a rather general ansatz for the curved string in
Sec.~\ref{ansaf}. Using this ansatz, we obtain analytic and numerical
solutions for the string shape in Secs.~\ref{solu_an} and
\ref{solu_num}. Finally, we sum up our results in Sec.~\ref{conclu}.

\section{Rotating string model with a nonzero relative time}
\label{model}

The RSM with a nonzero relative time has been studied in detail in
Ref.~\cite{Buis}. So, we simply recall here the main points of this
work. Starting
from the QCD Lagrangian and neglecting the spin contribution of the
quark and the antiquark, the Lagrange function of a meson reads
\cite{dubi94} ($\eta={\rm diag}(+---)$ and $\hbar=c=1$)
\begin{equation}
\label{nambu}
{\cal L}=-m_{1}\sqrt{\dot{\bm x}^{2}_{1}}-m_{2}\sqrt{\dot{\bm
x}^{2}_{2}}
-a\int^{1}_{0}d\theta\sqrt{(\dot{\bm w}\bm w')^{2}-\dot{\bm w}^{2}{\bm
w'}^{2}}.
\end{equation}
The first two terms are the kinetic energy operators of the quark and
the antiquark, whose current masses are $m_{1}$ and $m_{2}$. These two
particles are attached by a string with a tension $a$. $\bm x_{i}$ is
the coordinate of the quark $i$ and $\bm w$ is the coordinate of the
string. $\bm w$ depends on two variables defined on the string
world sheet: One is spacelike, $\theta$, and the other timelike, $\tau$.
Derivatives are denoted $\bm w'=\partial_{\theta}\bm w$ and
$\dot{\bm w}=\partial_{\tau} \bm w$. In this picture, $\tau$ is the
timelike evolution parameter for the string and the quarks. Let us
mention that bold quantities will always denote four-vectors.

Introducing auxiliary fields to get rid of the square roots in the
Lagrangian~(\ref{nambu}) and making the straight line ansatz to describe
the string, an effective Lagrangian can be derived \cite{guba94}
\begin{eqnarray}
\label{step2}
{\cal L}&=& -\frac{1}{2} \left[
\frac{m^{2}_{1}}{\mu_{1}}+\frac{m^{2}_{2}}{\mu_{2}}+
a_{1}
\dot{\bm R}^{2}+2a_{2}\dot{\bm R}\dot{\bm r}-2(c_{1}+\dot{\zeta}a_{1})
\dot{\bm R}\bm r\right.\nonumber\\
&&\left.-2( c_{2}+\dot{\zeta}a_{2})\dot{\bm r}\bm r
+a_{3}\dot{\bm r}^{2}+(a_{4}+2\dot{\zeta}c_{1}+\dot{
\zeta}^{2}a_{1})\bm r^{2}
\right],
\end{eqnarray}
where the coefficients $a_{1}$, $a_{2}$, \dots, are functions of the
auxiliary fields. Their exact expressions can be found in
Ref.~\cite{Buis}. The parameter $\zeta$ defines the position $\bm R$ of
the center of mass
\begin{equation}\label{cmdef}
\bm R=\zeta \bm x_1+(1-\zeta) \bm x_2\equiv (\bar{t},\vec{R}\, ),
\end{equation}
and $\bm r$ is the
relative coordinate
\begin{equation}\label{reldef}
\bm r=\bm x_{1}-\bm x_{2}\equiv (\sigma,\vec{r}\, ).
\end{equation}
It is worth mentioning that the auxiliary field $\mu_i$ can be
interpreted as the constituent mass of the quark whose current mass is
$m_i$. Moreover, the parameter $\zeta$ appears to be a function of the
auxiliary fields. It reduces to the expected value $\zeta=1/2$ when the
quark and the antiquark have the same mass \cite{sema04,buis042}. The
straight line ansatz for the string implies that the string coordinates
are given by
\begin{equation}\label{slans}
\bm w=\bm R+(\theta-\zeta)\bm r.
\end{equation}
Such an ansatz is suggested by lattice QCD calculations, which show that
the chromoelectric field between the quark and the antiquark appears to
be roughly constant on a straight line joining the two particles
\cite{Koma}.

The usual approach is to work with the equal time ansatz, i.~e.
\begin{equation}\label{tegal}
x^{0}_{1}=x^{0}_{2}=w^{0}=\tau=\bar{t}.
\end{equation}
Then, we have $\bm r=(0,\vec{r}\, )$, $\bm R=(t,\vec{R}\ )$,
$\dot{\bm r}=(0,\dot{\vec{r}}\, )$, and
$\dot{\bm R}=(1,\dot{\vec{R}})$. This procedure considerably simplifies
the equations, but neglects the relativistic retardation effects due to
a possible nonzero value of the relative time $\sigma$. That is why we
made in Ref.~\cite{Buis} a less restrictive hypothesis: We
identified the temporal coordinate of the center of mass with the
evolution parameter, $\bar{t}=\tau$, and we allowed a non vanishing
relative time $\sigma$. We have then
\begin{equation}\label{choice2}
\dot{\bm r}=(\dot{\sigma},\dot{\vec{r}}\, ),\ \dot{\bm
R}=(1,\dot{\vec{R}}\, ).
\end{equation}
It is then possible to derive from the Lagrangian~(\ref{step2}) a set of
three equations for the RSM with a nonzero relative time
\begin{subequations}
\label{seteq1}
\begin{eqnarray}
0&=&\mu_{1}y_{1}-\mu_{2}y_{2}-\frac{ar}{y_t}\left(\sqrt{1-y^{2}_{1}}-
\sqrt{1-y^{2}_{2}}\right), \label{peq0} \\
\frac{L}{r}&=&\frac{1}{y_t}(\mu_{1}y^{2}_{1}+\mu_{2}y^{2}_{2})+\frac{
ar} {y_t^{2}} (F(y_{1})+F(y_{2})), \label{lor} \\
H&=&\frac{1}{2}\left[\frac{p^{2}_{r}+m^{2}_{1}}{\mu_{1}}+\frac{p^{2}
_{r} +m^{2}_{2}}{\mu_{2}}+\mu_{1}(1+y^{2}_{1})+\mu_{2}(1+y^{2}_{2})
\right]
\nonumber \\
&&+\frac{ar}{y_t}(\arcsin y_{1}+\arcsin y_{2})+\Delta H, \label{haux}
\end{eqnarray}
\end{subequations}
with
\begin{equation}
\label{fy}
F(y_{i})=\frac{1}{2}\left[\arcsin y_{i}-y_{i}\sqrt{1-y^{2}_{i}}
\right] \quad \text{and} \quad y_t=y_{1}+y_{2}.
\end{equation}
$p_{r}$ is the radial momentum and $y_i$ is the transverse velocity of
the quark $i$. The first relation gives the cancellation of the total
momentum in the center of mass frame, while the two last ones define
respectively the angular momentum and the Hamiltonian of the system.
Equations~(\ref{peq0}) and (\ref{lor}) define the two variables $y_1$
and $y_2$. A further elimination of the auxiliary fields $\mu_1$ and
$\mu_2$ yields the relativistic flux tube model \cite{buis042}.

Equations~(\ref{seteq1}) are identical to
those of the usual RSM (see for example
Ref.~\cite{buis042}), but a perturbation of the Hamiltonian, denoted
$\Delta H$, is
now present. It contains the contribution of the retardation effects and
is given by \cite{Buis}
\begin{equation}\label{deltaH}
\Delta H=-\frac{\Sigma^{2}}{2a_{3}}+\frac{c_{2}}{a_{3}}\Sigma \sigma-(
c_{1}+\dot{\zeta}a_{1}) \sigma -\frac{c_{2}^{2}}{2a_{3}}\sigma^{2}+
\frac{1}{2}(a_{4}+2\dot{\zeta}c_{1}+\dot{\zeta}^{2}a_{1})\sigma^{2},
\end{equation}
where $\Sigma$ is the canonical momentum associated with the relative
time $\sigma$.

Let us now consider the quantized version of our model:
$L\rightarrow \sqrt{\ell(\ell+1)}$, $\left[r,\, p_{r}\right] =i$,
$\left[\sigma ,\, \Sigma\right] =-i$. The Hamiltonian~(\ref{haux}) has
then the following structure
\begin{equation}
H(\sigma,r)=H_{0}(r)+\Delta{H}(\sigma,r).
\end{equation}
The relative time only appears in the perturbation, and
$H_{0}$ depends only on the radius $r$. So, we can assume that the
total wave function reads
\begin{equation}
\left|\psi (\bm r)\right\rangle=\left |A(\sigma)\right\rangle\otimes
\left|R(r)\right\rangle\otimes\left|Y_{\ell\, m}(\theta,\phi)\right
\rangle,
\end{equation}
where $\left|R( r)\right\rangle$ is a solution of the
eigenequation
\begin{equation}\label{eigen1}
H_{0}(r)  \left|R(r)\right\rangle=M_{0}\left|R(r)\right\rangle.
\end{equation}
Such a problem can be solved for instance by the Lagrange mesh
technique \cite{buis041}. As it is shown in Ref.~\cite{Buis}, the total
mass is written
\begin{eqnarray}
M&=&M_{0}+\left\langle  A(\sigma) \right| \otimes\left\langle  R(r
) \right| \Delta H(r,\sigma)\left|R(r)\right\rangle\otimes\left
|A(\sigma) \right\rangle \nonumber\\
&=&M_{0}+\Delta M. \label{mtot}
\end{eqnarray}
The contribution $\Delta M$ is then given by the fundamental state of
the eigenequation
\begin{equation}\label{eigenret}
\Delta {\cal H}(\sigma)\left|A(\sigma)\right\rangle=\Delta M  \left|A(
\sigma)\right\rangle,
\end{equation}
where
\begin{equation}\label{average}
\Delta{\cal H}(\sigma)= \left\langle R(r) \right| \Delta
H(r,\sigma) \left|R(r)\right\rangle.
\end{equation}
In the case $m_1=m_2$, the eigenequation~(\ref{eigenret}) can
be solved, since $\Delta{\cal H}(\sigma)$ is simply a harmonic
oscillator in the relative time $\sigma$
\begin{equation}\label{dH4}
\Delta {\cal H}(\sigma)=-\frac{1}{2\langle a_{3}\rangle}\left[\Sigma^{2}
+
\langle c_{2}^{2}-a_{4}a_{3}\rangle\sigma^{2}\right].
\end{equation}
A more detailed study of the retardation term $\Delta M$ and of its
effect on the meson spectrum can be found in Ref.~\cite{Buis}. Let us
only observe two points from Eqs.~(\ref{eigenret}) and (\ref{dH4}).
Firstly, the retardation contribution $\Delta M$ is negative,
\begin{equation}
  \Delta M= -\frac{1}{2}\sqrt{\langle c_{2}^{2}-a_{4}a_{3}\rangle/
  \langle a_3\rangle^2},
\end{equation}
and it thus decreases the meson mass. Secondly, the temporal part of
the wave function reads
\begin{equation}\label{fotemp}
A(\sigma)=\left(\frac{\beta}{\pi}\right)^{1/4}\exp\left(-\frac{\beta}{2
}\sigma^{2}\right),
\end{equation}
with
\begin{equation}
\beta=\sqrt{\langle c_{2}^{2}-a_{4}a_{3}\rangle}.
\end{equation}
It is a gaussian function centered around $\sigma=0$. This provides an
interpretation of the equal time ansatz~(\ref{tegal}) as the most
probable configuration of the system.

\begin{figure}[ht]
  \begin{center}
    \includegraphics*[width=16cm]{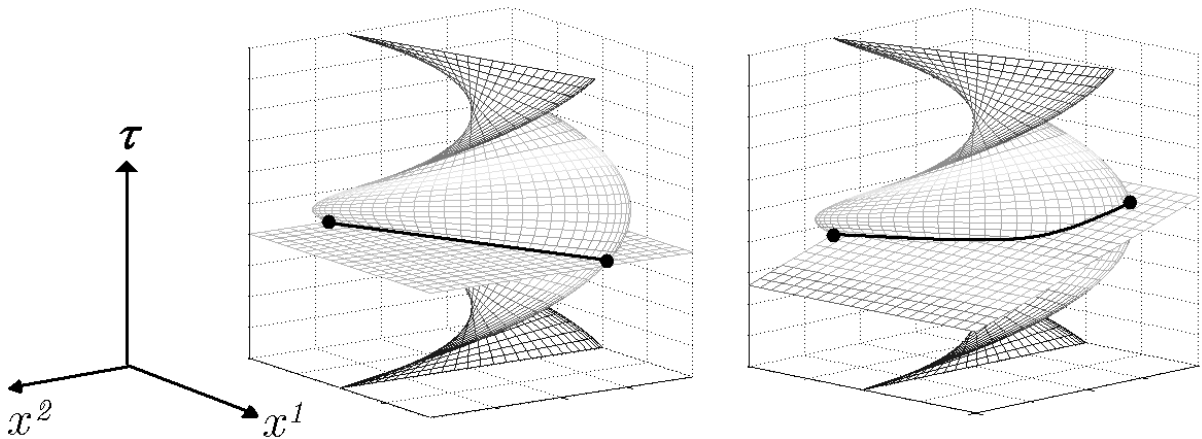}
  \end{center}
  \caption{Left: Intersection between a helicoidal string world sheet
  and
  a plane $\tau=\tau_0$ ($x^0_q(\tau_0)=x^0_{\bar{q}}(\tau_0)$), the
  intersection is a straight line. Right: Same situation, but now
  $x^0_q(\tau_0)\neq x^0_{\bar{q}}(\tau_0)$, and the intersection is
  curved.}
  \label{fig:helic}
\end{figure}

It is worth noting that the use of a non vanishing relative time is not
really compatible with the straight line ansatz. This can be seen by the
following simple considerations: Let us assume that the world sheet of
the system in the center of mass frame is a helicoid area in the case of
exactly circular quark orbits. The shape of the string is then a
straight line for a slice at constant time and a curve for a slice not
at constant time (see Fig.~\ref{fig:helic}). We already gave in
Ref.~\cite{Buis} a crude estimation of the possible bending of the
string due to retardation effects, and our result was compatible with a
small deviation from the straight line. The purpose of this paper is to
study more carefully the string deformations caused by the nonzero
relative time in order to confirm the validity of our approach.

\section{Constraints on the string shape}
\label{cons}

As the coordinates $\bm x_i$ are independent of the parameter $\theta$,
the Lagrangian~(\ref{nambu}) can be rewritten as
\begin{equation}
{\cal L}=\int^{1}_{0}d\theta{\cal \tilde{L}},
\end{equation}
with
\begin{equation}\label{lag1}
{\cal \tilde{L}}=-m_{1}\sqrt{\dot{\bm x}^{2}_{1}}-m_{2}\sqrt{\dot{\bm x}
^{2}_{2}}
-a\sqrt{(\dot{\bm w }\bm w')^{2}-\dot{\bm w }^{2}{\bm w'}^{2}}.
\end{equation}
Several steps are required to obtain a Hamiltonian from the
Lagrangian~(\ref{lag1}). Firstly, one has to find a particular solution,
denoted $\bm w^{*}$, for the string shape. This is in fact the solution
of the equations of motion (EOM) of the Nambu-Goto Lagrangian
\begin{equation}\label{lag2}
{\cal \tilde{L}}_{NG}=-a\sqrt{(\dot{\bm w }\bm w')^{2}-\dot{\bm w }^{2}{
\bm w'}^{2}},
\end{equation}
with the boundary conditions
\begin{equation}\label{cond}
\bm w(1,\tau)=\bm x_{1}(\tau),\ \ \bm w(0,\tau)=\bm x_{2}(\tau).
\end{equation}
Secondly, once $\bm w^{*}$ is known, it has to be injected in the total
Lagrangian~(\ref{lag1}). Thirdly, one has to compute the momenta defined
by
\begin{equation}\label{momen}
p_{i\mu}=\frac{\partial {\cal \tilde{L}}}{\partial \dot{x}^{\mu}_{i}}.
\end{equation}
In this picture, we assume that $\bm w^{*}(\theta,\tau)$ is in fact of
the form $\bm w^{*}(\theta,\bm{x}_i(\tau))$, because of the boundary
conditions~(\ref{cond}). Finally, with the momenta~(\ref{momen}), the
quantity
\begin{equation}
\label{ham2}
{\cal \tilde{H}}=\sum^{2}_{i=1}\bm p_i \dot{\bm x}_i-{\cal \tilde{L}}
\end{equation}
is readily computed, and the total Hamiltonian is thus given by
\begin{equation}
{\cal H}= \int^{1}_{0}d\theta{\cal \tilde{H}}.
\end{equation}
In the usual RSM, one uses the straight line ansatz~(\ref{slans})
together with the equal time ansatz~(\ref{tegal}). This provides indeed
a correct solution $\bm w^{*}$ of the EOM \cite[p. 122]{str}. The
question we want to answer in this section is: Is it possible to find a
more general form for the string shape?

Let us suppose that we have such a solution, $\bm w^{*}$, which
thus respects the conditions~(\ref{cond}). As we already remarked,
$\bm w^{*}$ depends on the coordinates $\dot{\bm x}_{i}(\tau)$. So, the
momenta~(\ref{momen}) read
\begin{equation}\label{momen2}
p_{i\mu}=-m_{i}\frac{\dot{x}_{i\mu}}{\sqrt{\dot{\bm{x}}^{2}_{i}}}-a
\frac{N_{\rho}}{\sqrt{\bm{N}\dot{\bm{w}}}}  \frac{\partial
\dot{w}^{\rho}}{\partial \dot{x}^{\mu}_{i}},
\end{equation}
with
\begin{equation}
\bm N=(\dot{\bm{w}}\bm{w}')\bm{w}'-\bm{w}'^{2}\dot{\bm{w}}.
\end{equation}
For the sake of simplicity, we have dropped the $^*$ on $\bm w$. The
Hamiltonian~(\ref{ham2}), computed with the momenta~(\ref{momen2}), is
expected to vanish since we work in a manifestly covariant formalism.
Here, we see that
\begin{equation}\label{ham3}
{\cal H}=a
\left[\sqrt{\bm{N}\dot{\bm{w}}}-\sum^{2}_{i=1}\frac{N_{\rho}}{\sqrt{\bm{
N}\dot{\bm{w}}}} \frac{\partial \dot{w}^{\rho}}{\partial \dot{x}^{\mu}_{
i}}\dot{x}^{\mu}_{i}\right].
\end{equation}
If we want ${\cal H}=0$, we must satisfy the condition
\begin{equation}\label{cons2}
\dot{w}^{\rho}= \sum^{2}_{i=1}\frac{\partial \dot{w}^{\rho}}{\partial
\dot{x}^{\mu}_{i}}\dot{x}^{\mu}_{i}.
\end{equation}
The solution for the constraint~(\ref{cons2}) is
\begin{equation}
\dot{w}^{\rho}(\theta,\tau)=  \sum^{2}_{i=1}A^{\rho}_{i\nu}(\theta,\tau)
\dot{x}^{\nu}_{i}(\tau),
\end{equation}
where the $A^{\rho}_{i\nu}(\theta,\tau)$ are unspecified functions. A
simple integration gives
\begin{equation}\label{form1}
w^{\rho}(\theta,\tau)=
\sum^{2}_{i=1}A^{\rho}_{i\nu}(\theta,\tau)x^{\nu}_{i}(\tau)-\sum^{2}_{i=
1}\int^\tau_0 d\tilde{\tau} \dot{A}^{\rho}_{i\nu}(\theta,\tilde{\tau})x^
{\nu}_{i}(\tilde{\tau})+f^{\rho}(\theta).
\end{equation}
Expressing Eq.~(\ref{form1}) in the center of mass and relative
coordinates, given by relations~(\ref{cmdef}) and (\ref{reldef}), we
obtain
\begin{eqnarray}\label{form2}
w^{\rho}&=&(A^{\rho}_{1\nu}+A^{\rho}_{2\nu}) R^{\nu}+\left[(1-\zeta)A^{
\rho}_{1\nu}-\zeta A^{\rho}_{2\nu}\right]r^{\nu}\\ \nonumber
&&-\int d\tilde{\tau}
\left\{(\dot{A}^{\rho}_{1\nu}+\dot{A}^{\rho}_{2\nu}) R^{\nu}+\left[(1-
\zeta)\dot{A}^{\rho}_{1\nu}-\zeta \dot{A}^{\rho}_{2\nu}\right]r^{\nu}
\right\} +g^\rho,
\end{eqnarray}
if we assume that $\dot{\zeta}=0$. We did not write explicitly the
dependences in $(\theta, \tau)$ to clarify the notations.

Formula~(\ref{form2}) can be simplified by taking the limit
$\bm{r}\rightarrow0$. In this case, we want indeed to obtain
$\bm{w}=\bm{R}$. This clearly implies that
\begin{equation}
A^{\rho}_{1\nu}+A^{\rho}_{2\nu}=\delta^{\rho}_{\nu},\ \ g^\rho=0,
\end{equation}
and the whole expression~(\ref{form2}) can thus be rewritten by using a
single set of functions $A^{\rho}_\nu\equiv A^{\rho}_{1\nu}$
\begin{equation}\label{solu1}
w^{\rho}=R^{\rho}+(A^{\rho}_{\nu}-\zeta\delta^{\rho}_{\nu})r^{\nu}-\int
d\tau \dot{A}^{\rho}_{\nu}r^{\nu}.
\end{equation}

The solution~(\ref{solu1}) is the general string shape we search for.
Solving the corresponding EOM would allow us to determine the functions
$A^{\rho}_\nu$. However, this problem is very complex, and we need to
make some simplifications to go further. We want to find a solution
which corresponds to a string configuration which does not change in
time. So we consider the functions $A^\rho_\nu$ to be independent of
$\tau$. We also make the following hypothesis
\begin{equation}
A^{\rho}_{\nu}=A^{\rho}\delta^{\rho}_{\nu},
\end{equation}
without summation on $\rho$. This implies that we want
$w^{\mu}$ depending only on the coordinates $R^{\mu}$ and $r^{\mu}$.
With our assumptions, the string has finally the following form
\begin{equation}\label{solu2}
w^{\mu}(\theta,\tau)=R^{\mu}(\tau)+\left[A^{\mu}(\theta)-\zeta\right]r^{
\mu}(\tau),
\end{equation}
without summation on $\mu$. The remaining unknown quantities are the
functions $A^{\mu}(\theta)$. Let us note that the straight string ansatz
is only a particular case of formula (\ref{solu2}), where
$A^\mu(\theta)=\theta$.

\section{The curved string ansatz}
\label{ansaf}

In order to further simplify the problem, we will here consider mesons
in which the quark and the antiquark have the same mass. In this case,
$\zeta=1/2$, and only the part of the string linking the center of mass
and one quark has to be known. The second part can be computed from the
first one by using symmetry arguments, as it will be discussed in
Sec.~\ref{solu_num}. Formally, we thus only have to find a solution for
$\bm w(\rho,\tau)$ with $\rho\in\left[0,1\right]$, such that
\begin{equation}\label{bord2}
\bm w(0,\tau)=\bm R,\ \ \bm w(1,\tau)=\bm r_q(\tau),
\end{equation}
$\bm r_q(\tau)$ being the position of the quark, for instance.
As we work in the center of mass frame, we can set
$\bm R=(\bar{t},\vec{0}\, )$. We can then rewrite Eq.~(\ref{solu2}) on
the following form
\begin{subequations}\label{form3}
\begin{eqnarray}
w^{0}(\rho,\tau)&=&\bar{t}(\tau)+A^{0}(\rho)r^{0}_q(\tau),\\
w^{i}(\rho,\tau)&=&A^{i}(\rho)r^{i}_q(\tau).
\end{eqnarray}
\end{subequations}

The meson evolves in a plane. We can thus set $A^3$ equal to zero
and use the complex coordinates $(w^{0},w,w^{*})$ defined by \cite{Olss}
\begin{equation}
w=\frac{1}{\sqrt{2}}(w^{1}+i w^{2}) .
\end{equation}
The Nambu-Goto Lagrangian is invariant for the reparameterization
of the world sheet \cite[p. 93]{str}, that is to say the
transformations
\begin{equation}
\tau'=\tau'(\rho,\tau),\ \ \rho'=\rho'(\rho,\tau).
\end{equation}
It allows us to fix for simplicity
\begin{equation}\label{rela}
\bar{t}(\tau)=\tau,\quad {\rm and}\quad A^1(\rho)=\rho.
\end{equation}
In the following, the relations~(\ref{rela}) will always be assumed. Let
us note that $\bar{t}(\tau)=\tau$ was a hypothesis of the model in
Ref.~\cite{Buis}.

It can be shown that, when $r^0_q(\tau)=0$,
\begin{subequations}\label{solustra}
\begin{eqnarray}
w^{0}(\rho,\tau)&=&\tau, \\
w(\rho,\tau)&=&\frac{r_q(\tau)}{\sqrt{2}}\, \rho\, \exp\left[i\omega
\tau \right],
\end{eqnarray}
\end{subequations}
is a solution of the EOM of the Nambu-Goto, or equivalently, of the
Polyakov Lagrangian (see for example Ref.~\cite{Olss}). In our
coordinates, these EOM read
\begin{subequations}\label{eom}
\begin{equation}
  \partial_a\left[\sqrt{-h}\, h^{ab} \partial_b w^\mu\right]=0,
\end{equation}
with $h^{ab}$ the inverse matrix of $h_{ab}$, given by
\begin{eqnarray}
  h_{ab}&=&\partial_a w^\mu \partial_b w_\mu ,\nonumber\\
  &=&-\partial_a w^0\partial_b w^0+2 {\rm Re}\left(\partial_a w^*
  \partial_b w\right).
\end{eqnarray}
\end{subequations}
The indices $(a,b)$ label the parameters $(\tau,\rho)$, and $h_{ab}$ is
the induced metric on the string world sheet.

Solution~(\ref{solustra}) describes a straight string with a constant
angular speed, and it corresponds to the following choices in
Eqs.~(\ref{form3}): $r^{0}_q(\tau)=0$, $A^{2}(\rho)=\rho$,
$r^{1}_q(\tau)=r_q(\tau) \cos \omega \tau$, and
$r^{2}_q(\tau)= r_q(\tau) \sin \omega t$.

In our previous work on the retardation effects, we used the
following ansatz \cite{Buis}
\begin{subequations}\label{anz1}
\begin{eqnarray}
w^{0}(\rho,\tau)&=&\tau+\rho\, \sigma_q(\tau), \\
w(\rho,\tau)&=&\frac{r_q(\tau)}{\sqrt{2}}\, \rho\exp(i\omega \tau),
\end{eqnarray}
\end{subequations}
that is to say $r^{0}_q(\tau)=\sigma_q(\tau)$,
$A^{0}(\rho)=A^{2}(\rho)=\rho$,
$r^{1}_q(\tau)= r_q(\tau) \cos \omega \tau$, and
$r^{2}_q(\tau)=r_q(\tau) \sin \omega t$. However, the string defined by
Eqs.~(\ref{anz1}) is not a solution of the EOM. Indeed, as
$\sigma_q(\tau)$ is assumed to be small (we showed that the retardation
can be treated as a perturbation), we can write the EOM at the first
order in $\sigma_q$. After some calculations, one can check
from relations (\ref{eom}) that $\sigma_q$ must satisfy the following
equation
\begin{equation}\label{eqret}
r^{2}_q\rho^{2}\ddot{\sigma}_q-2r_q\dot{r}_q\rho^2\dot{\sigma}_q+\left[
\left( r^{2}_q\omega^{2}-r_q\ddot{r}_q+2\dot{r}^2_q\right)\rho^{2}-2
\right]\sigma_q= 0.
\end{equation}
As $\sigma_q$ is only a function of $\tau$, Eq.~(\ref{eqret}) must be
satisfied for every value of $\rho$. In particular, when $\rho=0$, we
observe that $\sigma_q=0$ is the only possible solution. This confirms
that a straight string is not compatible with a nonzero relative time.

Finally, it appears quite natural that the string could be curved
because of the addition of two effects: The
rotation of the meson and the finiteness of the gluon speed. A curved
string can be described by the ansatz
\begin{subequations}
\label{anz2}
\begin{eqnarray}
w^{0}(\rho,\tau)&=&\tau+\rho\, \sigma_q(\tau), \\
w(\rho,\tau)&=&\frac{r_q(\tau)}{\sqrt{2}}\, \left[\rho+if(\rho)\right]
\exp \left[i(\omega \tau+\phi(\tau)\right],
\end{eqnarray}
\end{subequations}
where the spatial deformation $f(\rho)$ has been introduced as a
counterpart to the relative time $\sigma_q(\tau)$. Moreover, the
eventuality of a non constant rotation speed is taken into account
through the angular acceleration $\phi(\tau)$. If $\sigma_q=0$, our
ansatz reduces to the one of Ref.~\cite{Olss}, where the string
deformations due to angular acceleration are studied.
Equations~(\ref{anz2}) clearly describe a curved string, as it can be
seen by rewriting
it when $\tau=0$ with $\phi(0)=0$. Then we have simply $w^1\propto\rho$
and $w^2\propto f(\rho)$.

\section{Analytic solution}
\label{solu_an}

Finding an exact expression of $f(\rho)$ which satisfies the EOM of the
Nambu-Goto Lagrangian is a very complex problem, out of the scope of
this paper. The use of approximations appears necessary in order to deal
with workable equations. We have given in Ref.~\cite{Buis} some
arguments to show that the deformation of the string should be small.
Assuming that point, we will linearize the EOM in $f(\rho)$ and its
derivatives. The angular acceleration will also be considered as small,
as it is done in Ref.~\cite{Olss}. Following the hypothesis of this last
reference, we will only keep the linear terms in $\phi$ and its
derivatives. Terms like
$\phi f,\, \phi \sigma_q,\, \phi\dot{\sigma_q},\, \phi\dot{r}_q,\dots $
will thus also be neglected. Moreover, we consider that
$\ddot{r}_q,\, \ddot{\sigma_q}\approx0$. Consequently, our solution will
only be valid in the case of small radial excitations.

After a tedious algebra, we find that the EOM (\ref{eom}) are
satisfied by the curved string~(\ref{anz2}) if
\begin{equation}\label{equad}
\left(1-A_1\rho-A_2\rho^2+A_3\rho^3\right)\partial^2_\rho f(\rho)+(B_1-
B_2\rho)\left(\rho \partial_\rho f(\rho)-f(\rho)\right)=C+D\rho-E\rho^2,
\end{equation}
with
\begin{subequations}\label{def}
\begin{eqnarray}
A_1&=&\sigma_q\frac{\dot{r}_q}{r_q}-3\dot{\sigma_q},\\
A_2&=&\omega^2r^2_q+\dot{r}^2_q+2\sigma_q\dot{\sigma_q}\frac{\dot{r}_q}{
r_q}-3\dot{\sigma_q}^2,\\
A_3&=&\omega^2\sigma_q r_q\dot{r}_q+\sigma_q\frac{\dot{r}^3_q}{r_q}-
\omega^2\dot{\sigma_q}r^2_q-\dot{\sigma_q}\dot{r}^2_q-\sigma_q\dot{
\sigma_q}^2\frac{\dot{r}_q}{r_q}+\dot{\sigma_q}^3,\\
B_1&=&\omega^2 r^2_q+3\omega^2\sigma_q^2+2\sigma_q^2\frac{\dot{r}^2_q}{r
^2_q}-2\sigma_q\dot{\sigma_q}\frac{\dot{r}_q}{r_q},\\
B_2&=&2\omega^2\sigma_q r_q\dot{r}_q+2\sigma_q\frac{\dot{r}^3_q}{r_q}+
\omega^2\dot{\sigma_q} r^2_q -3\omega^2\dot{\sigma_q}\sigma_q^2-2\dot{
\sigma_q}\dot{r}^2_q+2\sigma_q^2\dot{\sigma_q}\frac{\dot{r}^2_q}{r^2_q}+
2\sigma_q\dot{\sigma_q}^2\frac{\dot{r}_q}{r_q},\\
C&=&2\sigma_q\omega,\\
D&=&2\omega\sigma_q\dot{\sigma_q}-2\omega\sigma_q^2\frac{\dot{r}_q}{r_q}
+r^2_q\ddot{\phi},\\
E&=& -\omega^3\sigma_q r^2_q+\omega^3\sigma_q^3-2\omega\sigma_q \dot{r}^
2_q+2\omega\sigma_q^3\frac{\dot{r}^2_q}{r_q}+2\omega\dot{\sigma_q}r_q
\dot{r_q}-2\omega\sigma_q^2\dot{\sigma_q}\frac{\dot{r}_q}{r_q},
\end{eqnarray}
\end{subequations}
and the boundary conditions
\begin{equation}\label{bord3}
f(0)=f(1)=0.
\end{equation}
Conditions~(\ref{bord3}) are in fact equivalent to the initial boundary
conditions~(\ref{bord2}).

Before performing a numerical resolution of the differential
equation~(\ref{equad}), we can find an approximate analytic solution by
developing $f(\rho)$ in powers of $\rho$,
\begin{equation}
f(\rho)=\sum^\infty_{n=0}a_n \rho^n.
\end{equation}
Keeping in this series only the terms which satisfy Eqs. (\ref{equad})
and (\ref{bord3}) at the second order in $\rho$, one can find that
\begin{equation}\label{solua}
f(\rho)=f_C(\rho)+f_D(\rho)+f_E(\rho),
\end{equation}
with
\begin{subequations}
\label{solua2}
\begin{eqnarray}
f_C(\rho)&=&-\frac{C}{2}\, \rho\, (1-\rho),\\
f_D(\rho)&=&-\frac{(D+CA_1)}{6}\ \rho\, (1-\rho^2),\\
f_E(\rho)&=&\frac{1}{12}\left[E-A_1(D+CA_1)-C(A_2-B_1/2)\right]\ \rho\ (
1-\rho^3).
\end{eqnarray}
\end{subequations}
The solution at the lowest order is $f_C(\rho)$. It is a parabola whose
maximum is reached in $\rho=1/2$. The next functions, $f_D(\rho)$ and
$f_E(\rho)$, shift slightly the maximum at a value $\rho<1/2$. A
graphical representation of the solution~(\ref{solua}) is given in the
next section. It is worth mentioning that in the case of a vanishing
angular momentum, $\omega=0$, the solution is trivially $f(\rho)=0$.
Even when the retardation is included, the string is straight when the
angular momentum is zero.

We can check that if $\sigma_q=0$, our solution reduces to
\begin{equation}
\left.f(\rho)\right|_{\sigma_q=0}=-\frac{r^2_q\ddot{\phi}}{6}\rho(1-\rho
^2),
\end{equation}
which is precisely the result of Ref.~\cite{Olss}. However, as we are
only interested in the retardation effects, we will take $\ddot\phi=0$
in the latter.

\section{Numerical solution}
\label{solu_num}

\subsection{Evaluation of the coefficients}

The previous section gave us a qualitative idea of the string shape.
But, to complete our analysis, we need an estimation of the magnitude of
the deformation. To do this, we have to compute numerically the
different coefficients~(\ref{def}), not in the classical framework we
used up to now, but in a quantized model. The coefficients (\ref{def})
should then be seen as operators whose average value has to be computed.
What should be done rigorously is to rewrite our RSM Hamiltonian with a
curved string solution of Eq.~(\ref{equad}) and compute the eigenstates
of this model to average the operators~(\ref{def}), in an analog way of
what is done in Ref.~\cite{Olss2}. But here, we only want to have a
first estimation of the string deformation. That is why, as we
considered that the deformation was small, it appears reasonable to
compute the mean values of the coefficients~(\ref{def}) with the states
of our particular RSM introduced in Ref.~\cite{Buis}. Let us remark
that,
if we want to be consistent with the notations for $r$ and $\sigma$ in
Sec.~\ref{model}, we have to make the following substitutions
\begin{equation}
r_q\rightarrow r/2,\quad\sigma_q\rightarrow\sigma/2,
\end{equation}
the angular speed $\omega$ being not modified. Our procedure will be to
replace each positive definite operator $X$ appearing in the
definitions~(\ref{def}) by
$\sqrt{\left\langle X^2\right\rangle}$, a quantity easy to compute with
our numerical method \cite{buis041}. This simple computation,
neglecting symmetrization problems, will only give us a rough estimation
of the different coefficients, but it is sufficient for our purpose.

We already mentioned that our numerical method, the Lagrange mesh
method, ensures us to know the radial wave function
$\left|R(r)\right\rangle$. The average values
$\left\langle \vec{p}\,^{2}\right\rangle=\left\langle
p^2_r+\ell(\ell+1)/r^2\right\rangle$ and $\left\langle r^2\right\rangle$
are easy to compute \cite{buis041}. Moreover, it is shown in Ref.~\cite
{buis05b} that the angular speed can be approximated by
\begin{equation}\label{omegadef}
\omega \approx\frac{\sqrt{\ell(\ell+1)}}{\left\langle r^2\right\rangle
\left(\left\langle \sqrt{\vec{p}\,^{2}+m^2}\right\rangle+a
\left\langle r\right\rangle/6\right)},
\end{equation}
where $\ell$ is the orbital angular momentum. The last spatial term we
need is $\langle \dot{r}^2\rangle$, which is the radial part of
$\left\langle \dot{\vec{r}}\,^{2} \right\rangle$. It can be computed,
from the Lagrangian~(\ref{step2}), that~\cite{buis042}
\begin{equation}
\left\langle  p^{\, 2}_r\right\rangle=\left\langle (a_3 \dot{r}-c_2 r)^2
\right\rangle.
\end{equation}
Using the fact that
$\left\langle c_2\right\rangle\approx0$ \cite{Buis}, we have
\begin{equation}
\left\langle \dot{r}^2\right\rangle\approx\frac{1}{\left\langle a_3
\right\rangle^2} \left[\left\langle p^{\, 2}_r\right\rangle-\left\langle
c^2_2\right\rangle\left\langle r^2\right\rangle\right].
\end{equation}

We turn now our attention to the terms involving the relative time.
With the temporal wave function given by Eq.~(\ref{fotemp}), it is
easily computed that
\begin{equation}
\sqrt{\left\langle \sigma^2\right\rangle}=\frac{1}{\sqrt{2\beta}}.
\end{equation}
$\langle \dot{\sigma}^2\rangle$ can be computed in analogy with
$\langle \dot{r}^2\rangle$. From the
Lagrangian~(\ref{step2}), it can be shown that~\cite{Buis}
\begin{equation}
\left\langle \Sigma^2\right\rangle=\left\langle \left(c_2 \sigma-a_3
\dot{\sigma}\right)^2\right\rangle.
\end{equation}
Since
\begin{equation}
\left\langle \Sigma^2\right\rangle=\frac{\beta}{2},
\end{equation}
we have
\begin{equation}\label{sdotdef}
\left\langle \dot{\sigma}^2\right\rangle\approx\frac{\beta}{2\left
\langle a_3\right\rangle^2}\left[1-\frac{\left\langle
c^2_2\right\rangle}{\beta^2}  \right].
\end{equation}
We are now able to compute, at least in first approximation, the
coefficients~(\ref{def}).

\subsection{Results}

The more interesting case is the meson formed of two massless quarks.
The deformation of the string is indeed expected to be maximal in this
case since the relativistic effects are the most important. Of course,
we will choose $\ell\neq0$ to observe a nonzero deformation. With $m=0$
and the standard value $a=0.2$ GeV$^2$, we can compute the needed
average values thanks to formulas~(\ref{omegadef}) to (\ref{sdotdef}).
They are given in Table~\ref{params} for three different states.

\begin{table}[h]
\protect\caption{Average values of quantities involved in the
coefficients~(\ref{def}),
for different states. For simplicity, the quantities of the form
$\sqrt{\langle X^2\rangle}$ appearing in the first column are denoted
as $X$. We recall that $r_q=r/2$ and $\sigma_q=\sigma/2$.}
\label{params}
\begin{ruledtabular}
\begin{tabular}{cccc}
(n+1)L         & 1P    & 1F     & 2P \\
\hline
$r_q$ (GeV$^{-1}$)\ &\ 2.526\ &\ 3.376\ &\ 3.411 \\
$\sigma_q$ (GeV$^{-1}$)\ &\ 1.093\ &\ 1.072\ &\ 1.122 \\
$\omega$ (GeV)       &\ 0.086\ &\ 0.087\ &\ 0.036\ \\
$\dot{r}_q$      &\ 0.459\ &\ 0.303\ &\ 0.592\ \\
$\dot{\sigma}_q$ &\ 0.342\ &\ 0.257\ &\ 0.252\ \\
\end{tabular}
\end{ruledtabular}
\end{table}

\begin{figure}[t]
  \begin{center}
    \includegraphics*[width=8cm]{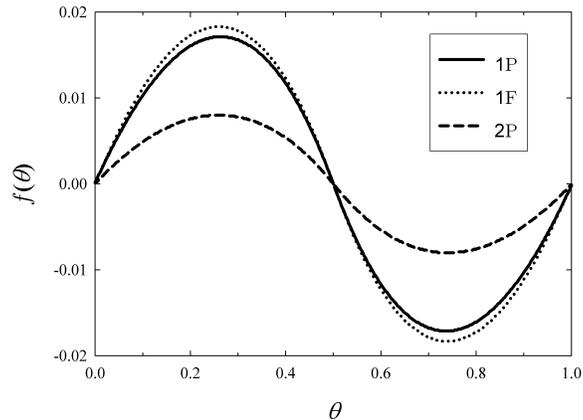}
  \end{center}
  \protect\caption{Numerical solution of Eq.~(\ref{equad}), giving the
  string shape between the quark and the antiquark for the states
  considered in Table~\ref{params}. $f(\theta)=0$ corresponds to the
  straight line ansatz.}
  \label{fig:d1}
\end{figure}

A numerical integration of Eq.~(\ref{equad}) to obtain the solution
$f(\rho)$ with the boundary
conditions (\ref{bord3}) can be performed by using the values of
Table~\ref{params}. As mentioned
in Sec.~\ref{ansaf}, one half of the string is simply given
by the couples $\left\{\rho,f(\rho)\right\}$. Once this solution is
known, we can compute every couple
$\left\{\theta, \tilde f(\theta)\right\}$
between the quark and the antiquark. Indeed, a simple symmetry argument
allows us to write
\begin{equation}\label{part}
\left\{\theta, \tilde f(\theta)\right\}=\left\{
\begin{array}{ll}
\left\{(1+\rho)/2,\, f(\rho)\right\} & \theta\in\left[1/2,1\right],\\
\left\{(1-\rho)/2,\, -f(\rho)\right\} & \theta\in\left[0,1/2\right].
\end{array}
\right.
\end{equation}
Relations~(\ref{part}) simply define a central symmetry with respect to
the center of mass.
The numerical solution of Eq.~(\ref{equad}) is plotted in
Fig.~\ref{fig:d1} for the three states considered in Table~\ref{params}.

We found that the deformation is maximal for the $1F$ state, and then
decreases when the quantum numbers increase. This is a consequence of
our approach, since we noticed in Ref.~\cite{Buis} that the retardation
effects are less large when the dynamical quark mass
$\langle \sqrt{\vec{p}\,^{2}+m^2}\rangle$ increases. In every case, the
deformation $f(\theta)$ is small with respect to one. This is an a
posteriori validation of our choice to consider small deformations only.

\begin{figure}[t]
  \begin{center}
    \includegraphics*[width=8cm]{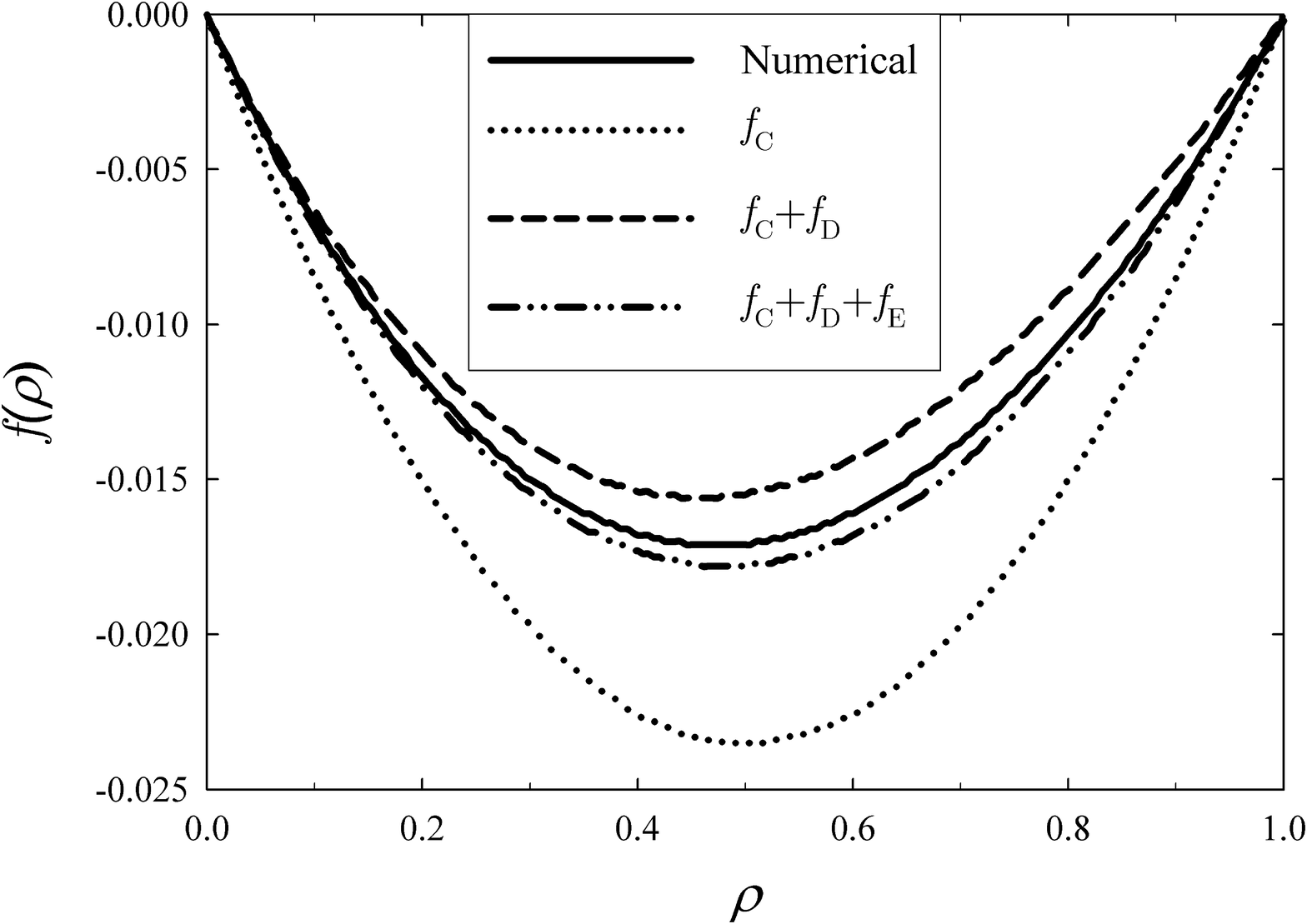}
  \end{center}
  \protect\caption{Comparison between the numerical solution of
  Eq.~(\ref{equad}) for the $1P$ state and the analytical
  formula~(\ref{solua}).}
  \label{fig:d2}
\end{figure}

In Fig.~\ref{fig:d2}, we compare the numerical solution for the $1P$
state, the lowest state in which the deformation occurs, with three
analytic approximations of this solution, given by
relations~(\ref{solua}) and (\ref{solua2}). It appears that $f_C$ is an
upper bound of the deformation, and that $f_C+f_D$ is sufficient to
correctly approximate the numerical solution.

Since the string brings an energetic contribution to the meson
which is proportional to its length, it is interesting to
evaluate the ratio between the lengths of both the curved and the
straight strings. It is given by
\begin{eqnarray}
\frac{\Delta L}{L}&=&\left[\int^1_0 d\rho \sqrt{1+(\partial_\rho f)^{2}}
\right]-1, \\
&\approx& \int^1_0 d\rho \frac{(\partial_\rho f)^{2}}{2}.
\end{eqnarray}
For a small deformation, as it is the case here, we can obtain an upper
bound for $\Delta L/L$ by using $f_C$ instead of $f$. To the second
order in $C$, we find that
\begin{equation}
\frac{\Delta L}{L}\leq\frac{C^2}{24},
\end{equation}
with $C=2\sigma_q\omega$. It is maximal in the $1F$ state. We
finally obtain the upper bound
\begin{equation}
\frac{\Delta L}{L}\leq 2\times 10^{-3}.
\end{equation}
The length of the string is only modified by some tenths of percent,
because of the bending induced by retardation effects. As the typical
mass scale for the mesons is the 1-2~GeV, the correction due to curved
string is around 1-4~MeV, as it is observed in Ref.~\cite{Olss2}. The
contribution of the bending of the string to the mass spectrum seems
thus very small. This is also small compared with the retardation
contribution, which can be around 100~MeV for massless quarks.

\section{Conclusion}\label{conclu}

We developed in this paper the idea that, if the retardation effects are
included in the rotating string model, the string linking the quark and
the antiquark cannot remain a straight line. We found a relation
constraining its shape, and obtained a general form for the string. The
straight line ansatz, which is valid in the equal time approximation,
appears to be only a particular case of our general form. When the
retardation is included, we showed that a bending of the string must be
taken into account, and we proposed a general ansatz defining a curved
string. With this ansatz, and assuming a priori that the deformation of
the string is small, we derived a differential equation giving its
shape. Analytical and numerical solutions of this equation can be found.
Roughly, the deformed string has a parabolic shape, with a small
amplitude. The amplitude of the deformation decreases when the quantum
numbers increase; but it is zero for a vanishing angular momentum. We
finally argued that the contribution of the bending of the string to the
mass spectrum is around 1-4~MeV. The typical meson mass scale is around
1-2~GeV, and the retardation contribution computed in a previous work
with a straight string is always smaller than 100~MeV. So, the effects
of the curvature are very weak, although rather interesting from a
theoretical point of view. However, if one is mainly interested in the
computation of a mass spectrum, we can conclude from our present work
that the rotating string with nonzero relative time can give
satisfactory results, event with a straight string approximation. We
leave the analysis of the rotating string model equations with nonzero
relative time and curved string for future work.

\acknowledgments

C.~S. (FNRS Research Associate) and F.~B. (FNRS Research
Fellow) thank the FNRS for financial support. V.~M. (IISN Scientific
Research Worker) thanks the IISN for financial support.

\end{document}